# Thermionic Emission as a tool to study transport in undoped nFinFETs

Giuseppe C. Tettamanzi, Abhijeet Paul, Gabriel P. Lansbergen, Jan Verduijn, Sunhee Lee, Nadine Collaert, Serge Biesemans, Gerhard Klimeck and Sven Rogge

**Abstract— Thermally activated sub-threshold transport has been investigated in *undoped* triple gate MOSFETs. The evolution of the barrier height and of the active cross-section area of the channel as a function of gate voltage has been determined. The results of our experiments and of the Tight Binding simulations we have developed are both in good agreement with previous analytical calculations, confirming the validity of thermionic approach to investigate transport in FETs. This method provides an important tool for the improvement of devices characteristics.**

## I. Introduction

In recent years, many MOSFET geometries have been introduced to overcome Short Channel Effects (SCE's) [1]. Among them, one of the most promising is the FinFET *geometry* [1-3]. In this structure, a much stronger gate-channel coupling can be obtained by the simultaneous action of the gate electrode on three faces of the channel (see Fig. 1a and 1b). The mechanisms of sub-threshold transport can be difficult to clarify due to the presence of screening [4]. As an example, the *undoped* channel version of these devices has a non-trivial and gate voltage ($V_g$) dependent current distribution. Therefore, the necessity of the development of tools that could be used to investigate current distribution has emerged. This knowledge is expected to allow an improvement of the device characteristics towards their scaling to the nanometers size regime. For device widths smaller than 5 nm, full volume inversion is expected to arise [1]. Wider devices are expected to be in the regime of *weak* volume inversion (where the bands in the channel closely follow the potential of $V_g$) for $V_g << V_{th}$ [1,5]. Several groups have investigated the behavior of such weak volume inversion devices using both classical [6,7], and quantum [8] computational models, but, to our knowledge, no experimental method that yields information on the location of the current-carrying regions of the channel exists. Taur has studied this problem analytically for an *undoped* channel with double gate (DG) geometry, using a 1-D Poisson equation [5]. The main conclusion emerging from this work is that, when the gate voltage is increased, a crossover takes place between the behavior of the channel at $V_g << V_{th}$, and at $V_g \sim V_{th}$, caused by screening of induced carriers which subsequently increase the carrier density at the gate-channel interface. To our knowledge, this prediction has never been directly observed experimentally. In our work we use a 2D model but the physical principle is fully analogues to the 1D case of Taur.

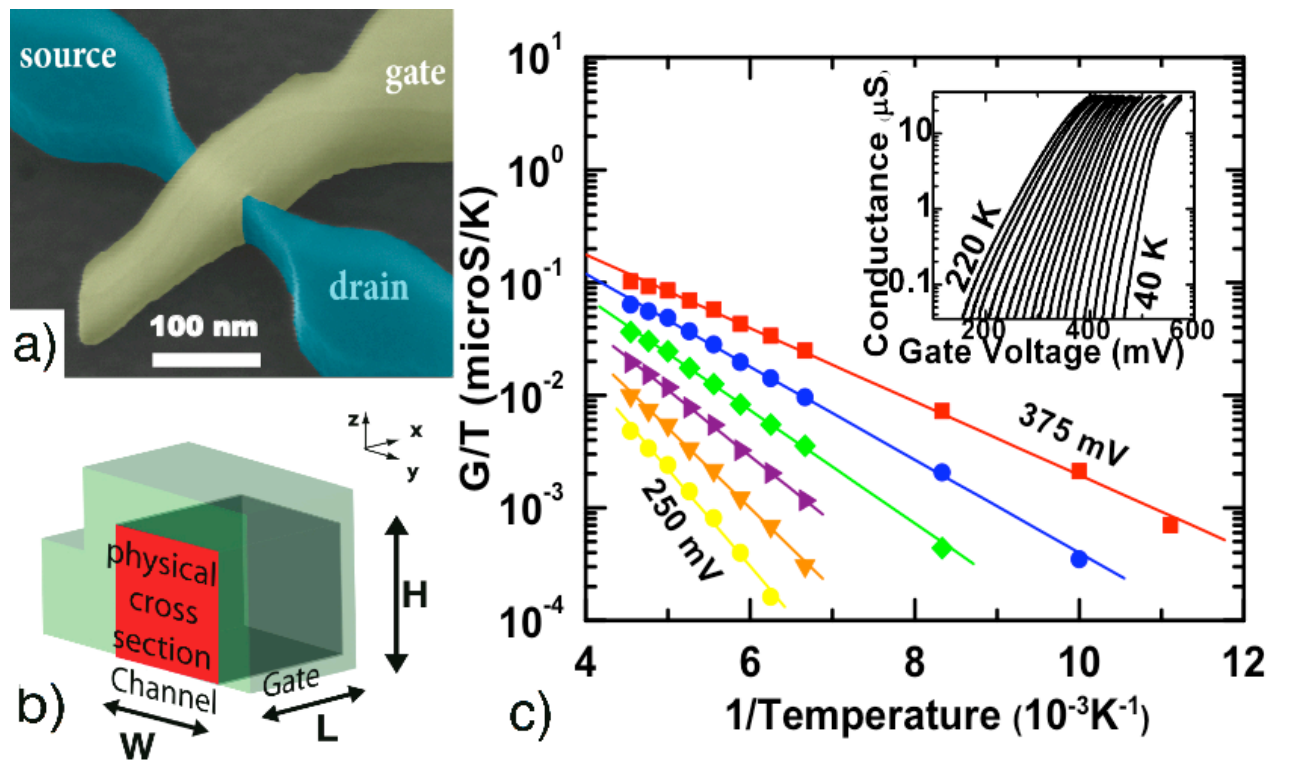

Fig. 1. a) Scanning Electron Microscope (SEM) image of a device. b) Schematic view of our FinFETs. The gate (light yellow) covers three faces of the channel (dark yellow). $L$, $H$ and $W$ represent the channel length, height and width respectively. In red the physical cross-sectional area is shown. c) Fits used to extrapolate $E_b$ and $S$ in one of our $W$=55 *nm* device. In the inset, differential conductance versus Vg data, for different temperatures, are shown.

## II. Experimental Results

Conductance versus temperature traces for a set of 8 *undoped* FinFET devices with the same channel length, ($L$=40 nm), and channel height, ($H$=65 nm), but different channel widths, ($W$= 25 nm, 55 nm, 125 nm and 875 nm) are presented. The discussion is focused on one device for each width since the same behavior for each of the devices of the same width is found consistently. Our devices consist of a nanowire channel etched on a 65 nm Si intrinsic film with a wrap-around gate covering three faces of the channel (Fig 1a and 1b) [3]. An HfSiO layer isolates a TiN layer from the intrinsic Si channel [3]. Differential conductance ($G=dI_{sd}/dV_{sd}$) data are taken at $V_{sd}$=0 $mV$ using a lock-in technique. Thermionic emission above a barrier can be represented by the formula [4,9]:

$$G = SA^{*}T\frac{e}{k_B}\exp(-\frac{E_b}{k_BT}) \qquad (1)$$

where $A^*=2.1 \times 120\ A\ cm^{-2}\ K^{-2}$ is the effective Richardson

Manuscript received September 2, 2009. The work of G.C.T., G.P.L., J.V. and S.R. and S. Rogge was supported by the FOM and the European Community's seventh Framework under the Grant Agreement nr:214989-AFSiD. The work of A.P., S.L. and G.K. was supported in part from SRC, MIND and MSD. G. Tettamanzi, G. Lansbergen, J. Verduijn and S. Rogge are with the Delft University of Technology, 2628 CJ Delft, The Netherlands (e-mail: G.C.Tettamanzi@tudelft.nl). A. Paul, S. Lee and G. Klimeck are with the Purdue University, West Lafayette, Indiana 47907, USA. N. Collaert and S. Biesemans are with IMEC, Kapeldreef 75, 3001 Leuven, Belgium.

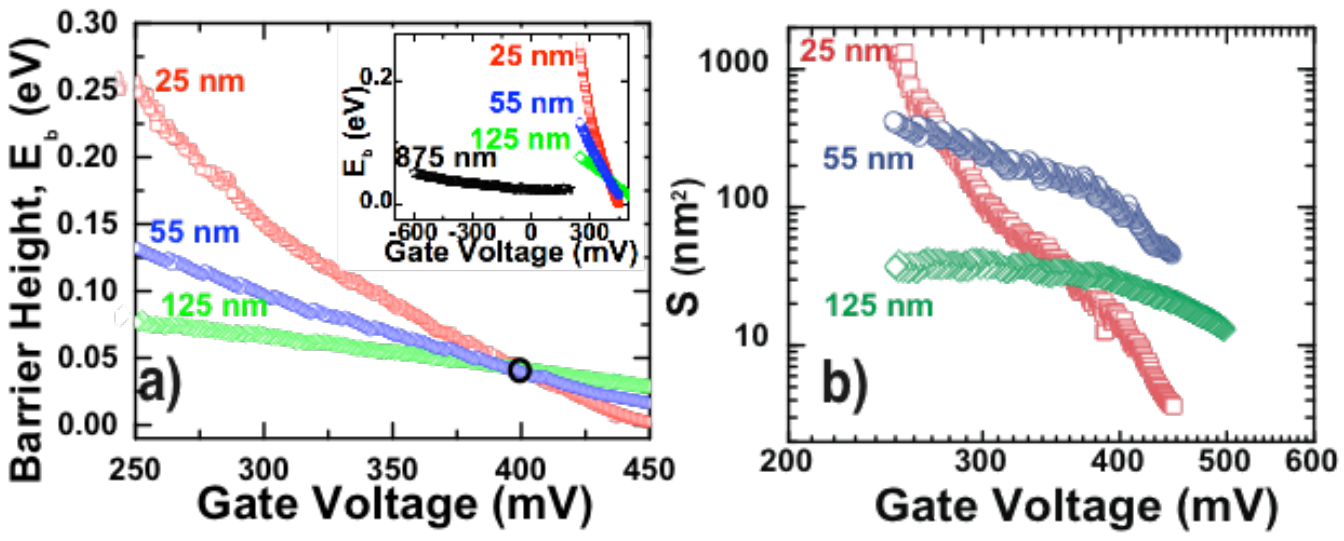

Fig. 2. Data obtained using the model of Eq. 1: a) $E_b$ versus $V_g$ for one device for each width from 25 nm to 125 nm. In the inset, calculated $E_b$ versus $V_g$ for all device widths are shown. b) Results of the dependence of the active cross section, *S*, versus $V_g$ obtained for all devices with $W \leq 125$ nm.

constant for Si [4,9]. Figure 1c shows the *G/T* versus *1000/T* data obtained from the re-arrangement of the *G* versus $V_g$ data taken at different temperatures (inset in 1c). Using the data of Fig. 1c, results for the source (drain) -channel barrier height, $E_b$, versus $V_g$ dependence and for the active cross-section area of the channel, *S,* versus $V_g$ dependence can be extrapolated using the thermionic fitting procedure. The important fact is that *S* is to be interpreted as a good estimation of the portion of the physical cross section area through which the transport preferentially occurs [4]. Note that Eq. 1 has only two parameters, *S* and $E_b$, and the accuracy obtained in the fits made using this equation ($R \sim 0.99$ for all fits of devices with width ≤ 125 nm as in the Fig. 1c) demonstrates the validity of the use of this model to study sub-threshold transport in our 40 nm channel length FinFETs. Fig. 2a examines the barrier height as a function $V_g$. We observe an expected decrease in $E_b$ while increasing $V_g$. As can be seen in the inset of Fig. 2a, this effect is less pronounced for wider device. We attribute this to short-channel effects (SCE's) that influence the electronic characteristics even at low bias. This trend is also reflected by the data of Table I, where the coupling factors obtained from our thermionic fits, $\alpha_1 = dE_b/dV_g$ [4] (thus the electrostatic coupling between the gate and the bulk of the channel), show a decrease for increasing width. In Table I, we also show the coupling between the potential of the channel interface and $V_g$, $\alpha_2$, extracted from Coulomb Blockade (CB) measurements (at 4.2 K) of confined states that are present at the Channel/Gate interface [4,10]. We find $\alpha_2$, to be a constant independent of *W*. In CB theory, $\alpha_2$ is the ratio between the electrochemical potential of the confined states and the change in $V_g$. This ratio can be estimated from the so called "stability diagram" [4]. This leads us to the conclusion that the coupling to the channel interface remains constant for increasing *W*, whereas the coupling to the center of the channel does not. In the 875 nm devices, SCE's are so strong (see inset Fig. 2a), that the thermionic theory loses accuracy; hence we will not discuss the results of these devices any further. All the $E_b$ versus $V_g$ curves, as depicted in Figure 2a, cross each other at around 0.4 V (outlined by the black circle), before complete inversion of the channel takes place at $V_{th} \sim 0.5$ V [3]. This suggests that at $V_g = 0.4$ V, the work function of the TiN is equal to the affinity of the Si channel in our devices [3]. The same value has also been verified in other measurements using capacitance-voltage (*C-V*) techniques [11], independently from the *W* of the channel. This fact gives us

TABLE I

Summary of the characteristics gate channel capacitive coupling of devices reported in this study, obtained from the results of Fit as in Fig 2a ($\alpha_1$) and from CB measurements at 4.2 K ($\alpha_2$).

| *Width (nm)* | $\alpha_1$ | $\alpha_2$ |
|---|---|---|
| 25 | 1 | 0.7 |
| *55* | 0.7 | 0.8 |
| *125* | 0.14 | 0.8 |
| *875* | 0.03 | 0.8 |

confidence that we are indeed observing activated transport over the channel barrier formed by the Metal/Oxide/Semiconductor interface, which at $V_g = 0.4$V will not dependent on *W*. The crossing point in Fig. 2a is not located exactly at $E_b = 0\ meV$, but is at *50 meV*. We attribute this feature to the presence, at the Channel-Gate boundary, of interface states (already found in CB measurements) that can store charge, repel electrons and therefore raise-up the barrier by a small amount. In $Si/SiO_2$ systems that have been studied in the past, these states were estimated to give an energy shift quantifiable between *70* and *120 meV* [10], in line with our data. Fig. 2b shows *S* as a function of $V_g$ extrapolated using Eq. 1. We can compare these results to the analytical model [5] discussed before and to our self-consistent simulations performed as described in [12-14]. At low $V_g$, devices with $W = 25\ nm$ show an active cross-sectional area of around 1000 $nm^2$ (see Fig. 2b). This is almost equal to the physical cross sectional area of the channel at these widths. At higher $V_g$, the active cross-sectional decreases to a few $nm^2$. We interpret this data as follows: at low $V_g$, transport in these devices is uniformly distributed everywhere in the physical cross-section of the channel (*weak* volume inversion), but with the increase of $V_g$, an increase of carriers density in the region near the interface, and, as a consequence, a reduction of *S,* arise. This interpretation corresponds with the screening mechanism discussed in [5]. Subsequently the action of the gate in the center of the channel is suppressed. Devices that have *55 nm* and *125 nm* widths behave in a fashion similar to the *25 nm* ones, but show a less pronounced decreasing trend and counter intuitive small values for *S,* as we observe a progressive reduction of $\alpha_1$ for increasing *W*. This is not a surprise as the barrier in these larger devices is lower and more carriers are allowed to migrate to the interface enhancing the screening effect. These results give for the first time an experimental insight into the mechanisms of conduction in *undoped* FinFETs.

## III. Comparison with Simulation

We used state-of-the-art-simulations, done using an atomistic 10 band sp3d5s* Tight-Binding (TB) model, to perform electronic structure calculation, coupled self-consistently with a 2D Poisson solver [12], and we obtained terminal characteristics using a ballistic top of the barrier model [13]. Due to the extensively large cross-section of our device that combines up to 44,192 atoms in the simulation domain, we had to integrate a new NEMO 3D code into our top of the barrier analysis [14]. With this expanded modeling capability we have been able to compare our experiment results with the simulations results. We do not expect effects of the variation of the potential in the source-drain direction to play a role in the device we simulated since $V_{sd}$ is very small [13,14]. Also, the gate length is long enough to suppress the tunneling current from source to drain [13,14]. In fact, using a geometry identical to the one of our FinFETs, with *W*= 25 nm, *H*= 65 nm and under biases similar to the ones of our experiments, the simulated current distribution shows a crossover from a situation of *weak* volume inversion at $V_g = 0$ *mV* (Fig. 3a) to a situation of transport confined prevalently at the interface at $V_g = 400$ *mV* (Fig. 3b). Therefore, also from the simulated current distribution, which gives a good indication of where mobile charges prevalently flow, a reduction of *S* with increasing $V_g$ can be extracted. However, this reduction is not as sharp as in the experimental data, as these simulations have been performed at *T*= 300 K and also due to the absence of interface states (expected to enhance the effect of screening in real devices) [4,10]. As a final benchmark to our experimental method, we have used the results of the TB simulations to calculate the current and the conductance at different temperatures and to extract, using again Eq. 1, simulated $E_b$ and *S* for a *W= 25 nm* device. In fact, in Figure 3c and 3d, we compare the simulated values with experiments and found that we can predict experimental results with good accuracy, although the simulations overestimate the values of *S* (probably for the same reasons discussed for Fig. 3b). In any case, by comparing our experimental results with the results of the simulations, we have a demonstration of the reliability of the method we have developed. This opens the way of its systematic use to obtain information about the magnitude and the position of carriers in FET devices in general and not only in our FinFET structures. In our investigations, we have neglected possible modifications of $A^*$ due to the constrained geometry [15] of our devices, as we found them to be negligible, and we have excluded tunneling regimes of transport [9,16] due to different temperatures dependences.

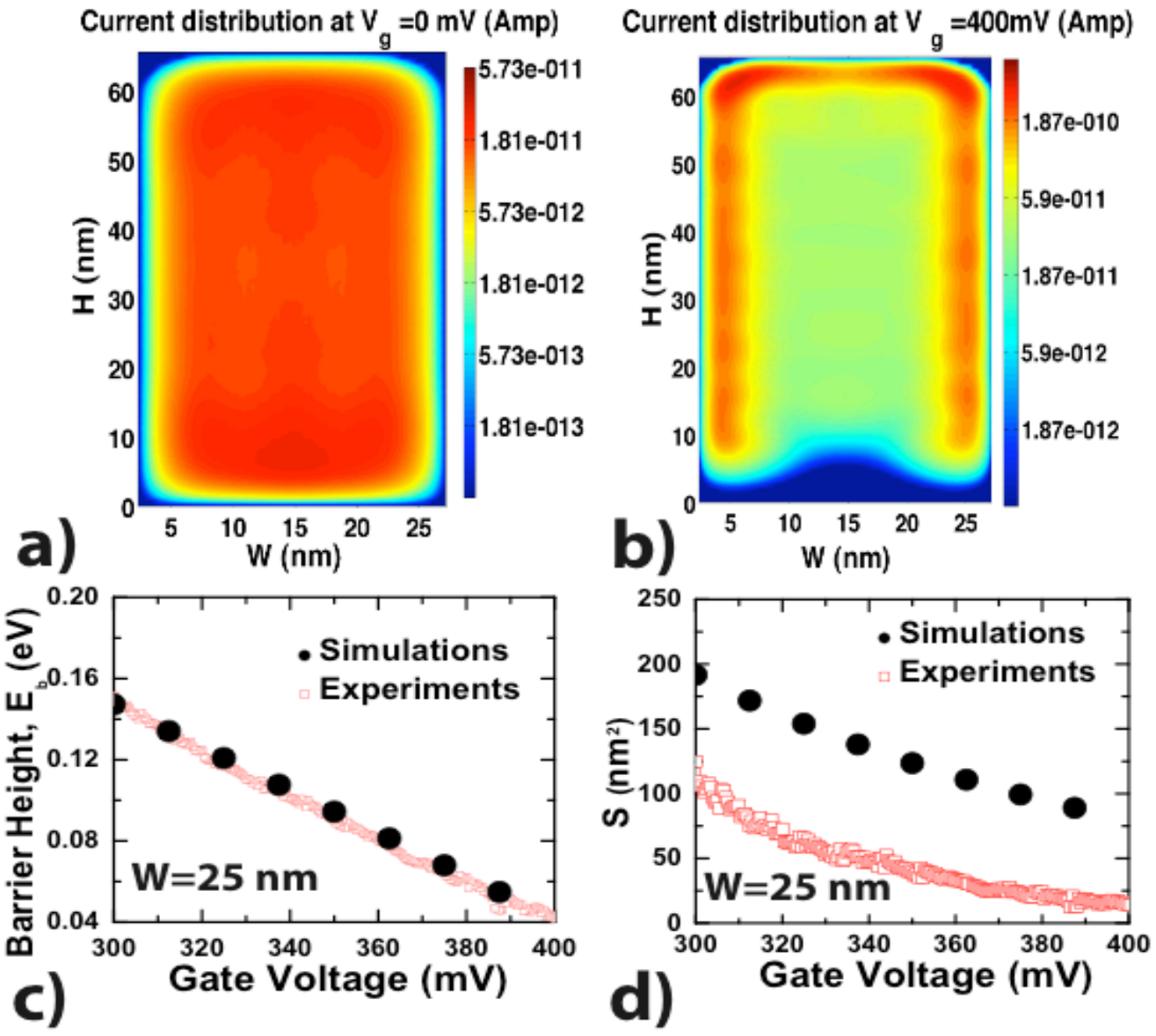

Fig. 3. Current distributions, for a) $V_g$ =0 *mV*, b) $V_g$ *=400 mV,* obtained using TB simulations for a FinFET geometry and *W= 25 nm*. Comparison of the simulated c) $E_b$ and d) *S* with the experimental data for the *W=25 nm device.*

## IV. Summary

In conclusion, we have presented results that are, at the best of our knowledge, the first experimental study of the behavior of the active cross-section area as a function of $V_g$ for *undoped* FinFETs. In particular, we present conductance traces for a set of *undoped* FinFET devices having the same channel length and height but different width, together with TB simulations for the device of *W=25 nm* and we compare them to theoretical calculations. For all these small devices (*W≤125 nm*), we propose a mechanism of inversion of the bands from flat band to band bending in the interface regions respectively, all as a function of $V_g$, therefore we have, for the first time, directly observed the theoretical result suggested by Taur [5]. By means of our results we have confirmed the validity of thermionic approach to investigate sub-threshold transport in FET devices and we have furthermore also given some answers to the fundamental technological question on how to localize and quantify areas of transport in these devices.

## Acknowledgment

The authors would like to thank P.A. Deosarran and J. Mol for useful discussions.